\documentclass[journal]{IEEEtran}
\usepackage{amsmath,amssymb,graphicx,dsfont,cite,color,bm,algorithm,algorithmic,epstopdf}
\usepackage{subfigure}
\usepackage{multirow}
\usepackage{multicol}
\begin{document}
\title{One-Bit MUSIC}
\author{Xiaodong Huang, and Bin Liao,~\IEEEmembership{Senior Member,~IEEE} \vspace{-2em} 
\thanks{The authors are with the Guangdong Key Laboratory of Intelligent Information Processing, College of Information Engineering, Shenzhen University, Shenzhen 518060, China (e-mail: binliao@szu.eud.cn).}
}

\markboth{ }%
{Shell \MakeLowercase{\textit{et al.}}: Bare Demo of IEEEtran.cls for IEEE Journals}

\maketitle

\begin{abstract}
In this letter, we consider the problem of direction-of-arrival (DOA) estimation with one-bit quantized array measurements. With analysis, it is shown that, under mild conditions the one-bit covariance matrix can be approximated by the sum of a scaled unquantized covariance matrix and a scaled identity matrix. Although the scaling parameters unknown because of the extreme quantization, they do not affect the subspace-based DOA estimators. Specifically, the signal and noise subspaces can be straightforwardly determined through the eigendecomposition of the one-bit covariance matrix, without pre-processing such as unquantized covariance matrix reconstruction. With so-obtained subspaces, the most classical multiple signal classification (MUSIC) technique can be applied to determine the signal DOAs. The resulting method is thus termed as one-bit MUSIC. Thanks to the simplicity of this method, it can be very easily implemented in practical applications, whereas the DOA estimation performance is comparable to the case with unquantized covariance matrix reconstruction, as demonstrated by various simulations.
\end{abstract}


\IEEEpeerreviewmaketitle

\section{Introduction}
\IEEEPARstart{A}{lthough} high-resolution quantization is preferred in terms of the signal recovery performance, it is likely to be impractical due to the high hardware cost and system power consumption, especially in emerging large-scale antenna array systems \cite{Sohrabi2018}. It is well known that the power consumption of an analog-to-digital converter (ADC) device is exponentially with quantization bit number \cite{Walden1999}. As a result, system design with low-resolution ADCs and related signal processing techniques have been attracted significant research interests over the past few years \cite{Singh2009,Wang2015,Mo2017}. Particularly, one-bit ADCs, which are composed of simple comparators, consume even negligible circuit power, and have been widely studied in massive multiple-input multiple-output (MIMO) \cite{Choi2016,Liang2016,Li2017}.

In this context, we herein consider the problem of direction-of-arrival (DOA) estimation with one-bit measurements, which is of great importance in both communication and radar. As a pioneer work, Bar-shalom and Weiss studied this problem in \cite{Bar-Shalom2002} and proposed to reconstruct the unquantized (original) covariance matrix according to the arcsine law \cite{Vleck1966,Jacovitti1994}. The reconstruction scheme is then employed for DOA estimation in sparse arrays such as nested and coprime arrays \cite{Liu2017, Chen2018}. On the other hand, with the development of one-bit compressive sensing (CS)\cite{Boufounos2008,Jacques2013}, the one-bit DOA estimation problem has been addressed through CS-based formulation \cite{Stockle2015, Yu2016, Huang2018}. By exploiting the signal sparsity in space domain, the one-bit DOA estimation problem is formulated as a sparse recovery problem in \cite{Stockle2015}. On the other hand, the signal sparsity in both space and frequency domains has been utilized in \cite{Yu2016}. More recently, a one-bit DOA estimator based on a fixed point continuation reconstruction algorithm is devised in \cite{Huang2018}.

Different from the above-mentioned existing methods which relies on unquantized (original) covariance matrix reconstruction or sparse signal recovery, we shall show in the sequel that the covariance matrix of one-bit array measurements, namely, one-bit covariance matrix, can be directly utilized to perform DOA estimation with subspace-based techniques. In specific, for uncorrelated signals with mildly low signal-to-noise ratio (SNR), it is proved that the one-bit covariance matrix can be approximated by a scaled unquantized covariance matrix with ignorable errors, except for the diagonal entries which are irrelevant to the signal and noise subspaces. Thus, subspace-based methods, such as multiple signal classification (MUSIC), can be straightforwardly applied without extra pre-processing. This leads to the so-called one-bit MUSIC approach. On this basis, the system hardware cost (including ADCs, data storage and transmission) and power consumption can be reduced, and the real-time implementation can be simplified. Both theoretical analysis and simulation results have verified the effectiveness of the one-bit MUSIC approach.

\section{One-Bit Signal Model}
Assume that $ K $ narrowband far-field signals impinge on an $ M $-element array from different directions $ \{ \theta_{1}, \theta_{2}, \cdots, \theta_{K} \}$. Under the assumption of infinite-resolution quantization, the output vector ${\bf x}(t)= \begin{bmatrix} x_{1}(t), \cdots, x_{M}(t) \end{bmatrix} ^ {T} \in {\mathbb C} ^ {M} $ of the array at time instant $ t $ can be expressed as
\begin{equation}
\label{1}
{\bf x}(t) = {\bf A}{\bf s}(t) + {\bf n}(t)
\end{equation}
where ${\bf A} = [ {\bf a}(\theta_{1}), \cdots, {\bf a}(\theta_{K}) ] \in {\mathbb C} ^{M \times K}$ represents the steering matrix with ${\bf a}(\theta)$ being the steering vector, ${\bf s}(t)= [ s_{1}(t), \cdots, s_{K}(t) ] ^ {T} \in {\mathbb C} ^ {K}$ and ${\bf n}(t)=[ n_{1}(t), \cdots, n_{M}(t) ] ^{T} \in {\mathbb C} ^{M}$ are, respectively, the signal vector and noise vector, and $ (\cdot)^{T} $ denotes the transpose. Note that the signal and noise are assumed to be uncorrelated, and both of them are modeled as independent, zero-mean, circular, complex, Gaussian random processes.

When one-bit ADCs are employed for quantization, the array output should be modified as
\begin{equation}
\label{2}
{\bf y}(t) = {\mathcal Q}( {\bf x}(t)) = {\mathcal Q}({\bf A}{\bf s}(t) + {\bf n}(t) )
\end{equation}
where ${\mathcal Q}(\cdot)$ represents a complex-valued element-wise quantization function composed of two sign functions ${\rm sign}(\cdot)$ as
\begin{equation}
\label{3}
{\mathcal Q}(z) = \frac{1}{\sqrt{2}} \left({\rm sign}(\Re\{z\}) + j {\rm sign}(\Im \{ z\})\right)
\end{equation}
where $\Re\{ z\}$ and $\Im\{ z\}$ denote the real part and imaginary part of a complex-valued number $z$, respectively.  

As a pioneer work for one-bit DOA estimation, Bar-shalom and Weiss \cite{Bar-Shalom2002} proposed to reconstruct the unqunatized covariance matrix ${{\bf R}}_{\bf x} = E[{\bf x}(t){\bf x}^H(t)]$ from the one-bit covariance matrix ${{\bf R}}_{\bf y} = E[{\bf y}(t){\bf y}^H(t)]$ by making use of the arcsine law \cite{Vleck1966}. Unlike this approach, we shall show that the subspace-based methods can be directly applied to the one-bit covariance matrix  $\bf R_y$ without extra pre-processing.

\section{One-Bit MUSIC}
To begin with, let us express the unquantized measurement of the $m$th sensor as
\begin{equation}\label{4}
  x_m(t) = \sum_{i=1}^K a_m(\theta_i)s_i(t) + n_m(t)
\end{equation}
where $ a_m(\theta_i)$ denotes the $m$th entry of ${\bf a}(\theta_i)$. Under the signal model described in Section I, it is known that $x_m(t)$ has zero mean and its variance $\sigma_{x_m}^2 =E[x_m(t)x_m^*(t)]$ (which is equal to the $m$th diagonal entry of $\bf R_x$) is given by
\begin{equation}\label{5}
 \sigma_{x_m}^2 =  [{\bf R_x}]_{mm} = \sum_{i=1}^K |a_m(\theta_i)|^2 \sigma_i^2 + \sigma_n^2
\end{equation}
where $\sigma_i^2$ and $\sigma_n^2$ denote powers of the $i$th signal and noise, respectively, see also \cite{Liao2016}. On this basis and recalling the independence of the signals and noise, the correlation coefficient between the unquantized measurements of the $m$th and $n$th sensors (where $m\neq n$) is given by
\begin{equation}\label{6}
  \begin{split}
    \rho_{x_mx_n} &=\frac{E[x_m(t)x_n^*(t)]}{\sigma_{x_m}\sigma_{x_n}}  = \frac{[{\bf R_x}]_{mn}} {\sqrt{[{\bf R_x}]_{mm}}\sqrt{[{\bf R_x}]_{nn}}} \\
                & = \frac{\sum_{i=1}^K a_m(\theta_i)a_n^*(\theta_i) \sigma_i^2}    {\sqrt{\sum_{i=1}^K |a_m(\theta_i)|^2 \sigma_i^2 + \sigma_n^2} \sqrt{\sum_{i=1}^K |a_n(\theta_i)|^2 \sigma_i^2 + \sigma_n^2}}\\
                & = \frac{\sum_{i=1}^K a_m(\theta_i)a_n^*(\theta_i) {\xi}_i}    {\sqrt{\sum_{i=1}^K |a_m(\theta_i)|^2 {\xi}_i+ 1} \sqrt{\sum_{i=1}^K |a_n(\theta_i)|^2 {\xi}_i + 1}}
  \end{split}
\end{equation}
where ${\xi}_i$ defines the SNR of the $i$the signal as
\begin{equation}\label{7}
  {\xi}_i = \frac{\sigma_i^2}{\sigma_n^2}.
\end{equation}

In order to simplify the analysis, we assume that the sensors are identical and isotropic, i.e., $a_m(\theta) = e^{j2\pi f \tau_m(\theta)}$ with $f$ being the carrier frequency and $\tau_m(\theta)$ being the time delay related to the location of the $m$th sensor. Thus, \eqref{5} is reduced to
$p\triangleq[{\bf R_x}]_{mm} = \sum_{i=1}^K  \sigma_i^2 + \sigma_n^2, ~m=1,\cdots,M.$
Furthermore, we assume that the signals are equal-power, i.e., $\xi_1=\cdots=\xi_k=\xi$, then we have
\begin{equation}
 \rho_{x_mx_n}  = \frac{\sum_{i=1}^K e^{j2\pi f (\tau_m(\theta_i)-\tau_n(\theta_i))}}{K + \xi^{-1}}.
\end{equation}
Obviously, it is known that $|\rho_{x_mx_n} |<1$. More importantly, both the real and imaginary parts of the correlation coefficient $\rho_{x_mx_n}$ are decreasing with the decrease of the signal SNR $\xi$. For illustration, we assume two signals from $5^\circ$ and $15^\circ$ impinge on a 10-element unform linear array (ULA) with half-wavelength inter-element spacing. The resulting curves of the real part $\Re\{\rho_{x_mx_n}\}$ (in red color) and imaginary part $\Im\{\rho_{x_mx_n}\}$ (in blue color) of the correlation coefficient versus SNR are depicted in Fig. 1.

\begin{figure}[!t]
  \centering
  \includegraphics[width=3.5 in]{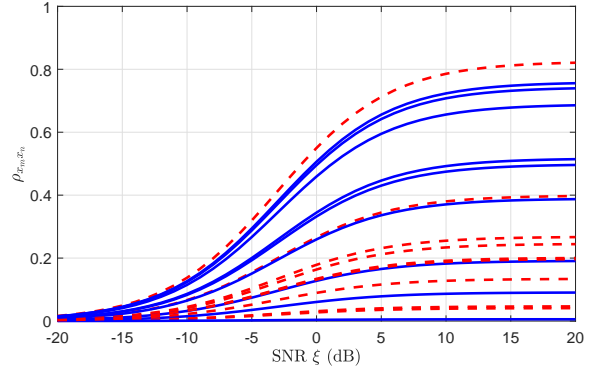}\\
  \caption{Correlation coefficient $\rho_{x_mx_n}$ versus SNR $\xi$.}\vspace{-1em}\label{Fig1}
\end{figure}

To proceed, we consider the one-bit quantized measurement $y_m(t)=\mathcal{Q}(x_m(t))$. According to the signal model, it is known that $y_m(t)$ has zero mean and unit variance, i.e., $E[y_m(t)]=0$ and $\sigma_{y_m}^2=1$, for $m=1,\cdots, M$. As a consequence, the correlation coefficient between the one-bit measurements $y_m(t)=\mathcal{Q}(x_m(t))$ and $y_n(t)=\mathcal{Q}(x_n(t))$ as
\begin{equation}\label{9}
  \rho_{y_my_n} =\frac{1}{\sigma_{y_m}\sigma_{y_n}} E[y_m(t)y_n^*(t)] = [{\bf R_y}]_{mn}.
\end{equation}
In particular, if $m=n$, we have $\rho_{y_my_m}=[{\bf R_y}]_{mm}= 1$. Furthermore, according to the arcsine law \cite{Vleck1966,Jacovitti1994}, we have
\begin{align}
\rho_{y_my_n}\! & = \frac{2}{\pi}{\rm arcsine}(\rho_{x_mx_n})\nonumber\\
            & \triangleq \frac{2}{\pi} \left( {\rm arcsin} \left( \Re\{\rho_{x_mx_n}\}\! \right) \!+\! j  {\rm arcsin} \left( \Im\{\rho_{x_mx_n}\}\! \right) \right)
\end{align}
which implies that $[{\bf R_y}]_{mn} = \frac{2}{\pi}{\rm arcsine}\left(\frac{1}{p}[{\bf R_x}]_{mn} \right)$, or
\begin{equation}\label{11}
 {\bf R_y} = \frac{2}{\pi}{\rm arcsine}\left(\frac{1}{p}{\bf R_x}\right).
\end{equation}
This means that one can reconstruct the unquantized covariance matrix from the one-bit covariance matrix as ${\bf R_x} = p^{-1}{\rm sine}\left(\frac{\pi}{2}{\bf R_y}\right)$, where ${\rm sine}(z) \triangleq \sin(\Re\{z\})+j \sin(\Im\{z\})$, and the unknown scaling parameter $p^{-1}$ which does not affect DOA estimation. Alternatively, we shall prove that ${\bf R_x} $ can be approximated by the summation of a scaled $\bf R_y$ and a scaled identity matrix $\bf I$.

\begin{figure}[!t]
  \centering
  \includegraphics[width=3.5 in]{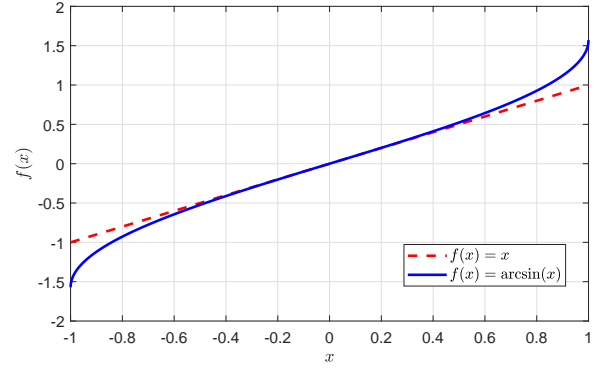}\\
  \caption{Comparison of $f(x)=x$ and $f(x)=\arcsin(x)$ for $x\in(-1,1)$.}\vspace{-1em}\label{Fig2}
\end{figure}

It has been shown above that $\Re^2\{\rho_{x_mx_n}\}+\Im^2\{\rho_{x_mx_n}\}<1$, $|\Re\{\rho_{x_mx_n}\}|<1$  and $|\Re\{\rho_{x_mx_n}\}|<1$ for $m\neq n$, and thus, ${\rm arcsin} \left( \Re\{\rho_{x_mx_n}\} \right) $ can be expanded as
\begin{align}\label{12}
    {\rm arcsin} \left( \Re\{\rho_{x_mx_n}\} \right) =~& \Re\{\rho_{x_mx_n}\}+\frac{1}{6}\Re^3\{\rho_{x_mx_n}\}\nonumber\\
    &+\frac{3}{40}\Re^5\{\rho_{x_mx_n}\}+\cdots
\end{align}
Obviously, if $|\Re\{\rho_{x_mx_n}\}|$ is small enough, or equivalently, the SNR ($\xi$) is sufficiently low, ${\rm arcsin} \left( \Re\{\rho_{x_mx_n}\} \right) $ can be well approximated by $\Re\{\rho_{x_mx_n}\}$, i.e., ${\rm arcsin} \left( \Re\{\rho_{x_mx_n}\} \right) \doteq \Re\{\rho_{x_mx_n}\}$, by ignoring high-order terms. Such an approximation can also be applied to $\Im\{\rho_{x_mx_n}\}$. To have a better understanding of this approximation, the function $f(x)=\arcsin(x)$ and its approximation $f(x)=x$ for  $x\in(-1,1)$ are compared in Fig. \ref{Fig2}. It is seen that except for $|x|=1$, the approximation can be well guaranteed.

Based on the above essential concept, it can be concluded that if the SNR is sufficiently low, we have $\rho_{y_my_n} \doteq \frac{2}{\pi}\rho_{x_mx_n}$. Thus, recalling \eqref{6}, \eqref{9} and \eqref{12}, we have
\begin{equation}\label{18}
  [{\bf R_y}]_{mn} \doteq \frac{2}{p\pi}[{\bf R_x}]_{mn}, ~m\neq n. 
\end{equation}
It should be pointed out that the above approximation cannot be well shared by the case of $m=n$, owing to the fact that for any SNR we have $\rho_{y_my_m}=\rho_{x_mx_m} =1$ and the error caused by approximating $\arcsin(1)$ to $1$ is relatively large, as shown in Fig. \ref{Fig2}.  Nevertheless, we can rewrite \eqref{18} as
\begin{equation}
  {\bf R_y} - {\mathcal D}({\bf R_y}) \doteq \frac{2}{p\pi}\left({\bf R_x} -  {\mathcal D}({\bf R_x}) \right)
\end{equation}
where ${\mathcal D}({\bf X}) = {\rm diag}\{[{\bf X}]_{11},\cdots, [{\bf X}]_{MM}\}$ returns a diagonal matrix. Because $[{\bf R_y}]_{mm}=1$ and $[{\bf R_x}]_{mm}=p$, we have
\begin{equation}\label{15}
  {\bf R_y}   \doteq \frac{2}{p\pi}{\bf R_x} + \left( 1- \frac{2}{\pi} \right){\bf I} \triangleq {\bf R}_{\bf y}^{\rm app}
\end{equation}
which implies that ${\bf R}_{\bf y}^{\rm app}$ and ${\bf R_x}$ have nearly identical eigenvectors, even though $p$ is unknown. As a consequence, the following essential proposition can be obtained.

{\emph{Proposition 1: For an array with identical and isotropic sensors, the signal and noise subspaces can be straightforwardly obtained from the eigendecomposition of one-bit covariance matrix, if the signal SNRs are sufficiently small.}}

According to the above proposition, it is known that classic subspace-based DOA estimation approaches, such as MUSIC, can be applied by performing the eigendecomposition of $\bf R_y$ to obtain the signal and noise subspaces. In practice, $\bf R_y$ can be estimated as $ {\widehat{\bf R}}_{\bf y} = \frac{1}{N}{\bf Y}{\bf Y}^H$, where $ {\bf Y} = \begin{bmatrix} {\bf y}(t_{1}), \cdots, {\bf y}(t_{N}) \end{bmatrix} $ include $N$ one-bit measurement vectors. It is worth mentioning that although theoretically the SNR should be small, extensive experiments show that subspace-based method using ${\bf R}_{\bf y}^{\rm app}$ can also provide good performance at high SNR levels.

Finally, the main steps of one-bit MUSIC for DOA estimation is summarized in Algorithm 1.
\begin{algorithm}[!h]
 	\caption{One-Bit MUSIC}
 	\label{alg1}
 	\begin{algorithmic}[1]
 		\STATE Collect one-bit measurements ${\bf Y}=[{\bf y}(t_1),\cdots, {\bf y}(t_N)]$.
 		\STATE Estimate the one-bit covariance matrix as ${\widehat{\bf R}_{\bf y}}=\frac{1}{N}{\bf YY}^H$.
        \STATE Perform eigendecomposition ${\widehat{\bf R}_{\bf y}}={\widehat{\bf U}}{\widehat{\bf\Sigma}}{\widehat{\bf U}}^H$.
        \STATE Output the estimate of noise subspace ${\widehat{\bf U}}_n$, i.e., the eigenvectors associated with the $M-L$ smallest eigenvalues.
        \STATE Carry out spectrum search $G(\theta) = \frac{1}{({\bf a}^H(\theta){\widehat{\bf U}}_n{\widehat{\bf U}}_n^H ){\bf a}(\theta)}$.
        \STATE Output the DOA estimates according to the locations of the $L$ highest peaks of the spectrum.
 	\end{algorithmic}
\end{algorithm}
\vspace{-1em}

\section{Simulation Results}
\begin{figure}[!t]
  \centering
  \includegraphics[width=3.5 in]{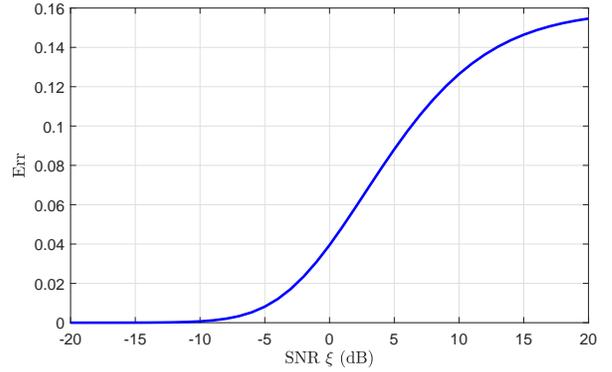}\\
  \caption{Approximation error of the one-bit covariance matrix ${\bf R}_{\bf y}^{\rm app}$.}\vspace{-1em}\label{Fig3}
\end{figure}

In this section, a set of simulations are carried out to verify the previous analysis and validate the effectiveness of one-bit MUSIC algorithm, and compare it with one-bit MUSIC with covariance matrix reconstruction using \eqref{11} and MUSIC with unquantized measurements. A 10-element ULA is considered and two equal-power narrowband signals impinge on the array from $-10^\circ$ and $3.5^\circ$. Both the signals and noise are drawn from independent and identically distributed complex Gaussian processes with zero mean and their variances are $ \rho_{s}^{2}$ and $ \rho_{n}^{2} $, respectively. The SNR is defined as $ {\rm SNR} = 10 \log_{10} \xi $ (in dB). Without loss of generality, we assume $\sigma_n^2=1$ in all simulations. The one-bit MUSIC algorithm using ${\bf R}_{\bf y}^{\rm app}$ is compared to the algorithm with covariance matrix reconstruction using $\bf R_y$(denoted as Recon. One-bit MUSIC)  and  the MUSIC algorithm without quantization (denoted as Unquantized MUSIC)  ,

At first, we examine the approximation error of ${\bf R}_{\bf y}^{\rm app}$. To this end, the error is defined as ${\rm Err} = \frac{\|{\bf R}_{\bf y}^{\rm app} - {\bf R_y}\|_F}{\|{\bf R_y}\|_F}$, where ${\bf R}_{\bf y}^{\rm app}$ and $\bf R_y$ are computed with \eqref{15} and \eqref{11}, respectively, by assuming that $\bf R_x$ and $p$ are known. The resulting approximation error versus SNR is depicted in Fig. \ref{Fig3}. It is seen that the error increases with the increase of SNR, which coincides with our previous analysis.

\begin{figure}
  \centering
  \includegraphics[width=3.5 in]{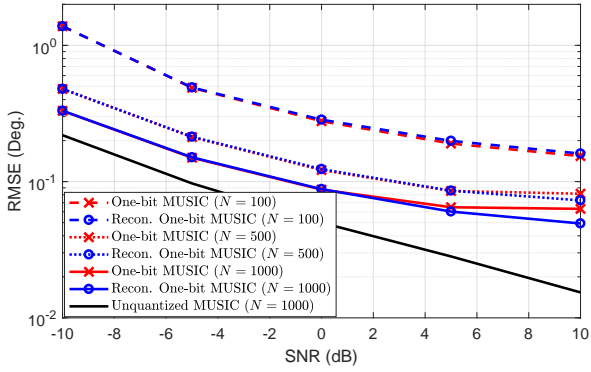}\\
  \caption{RMSE of DOA estimation versus SNR.}\vspace{-1em}\label{Fig4}
\end{figure}

In the second example, we examine the DOA estimation performance in terms of the root mean square error (RMSE) as ${\rm RMSE} = \sqrt{ \frac{1}{RK} \sum_{r=1}^{R} \sum_{i=1}^{K} ( \hat{\theta}_{i,r} - \theta_{i} )^{2} }$, where $ \hat{\theta}_{i,r} $ is the $i$th DOA estimate in the $r$th run and $ R = 1000 $ is the total number of Monte Carlo runs. The curves of RMSE versus SNR are shown in Fig \ref{Fig4}. For comparison, three different numbers of snapshots, i.e., $N=100$, $500$ and $1000$, are tested. It is seen that when the SNR is less than $0$ dB, there is no visible performance difference between the One-bit MUSIC and Reconstruction One-bit MUSIC. The performance gap becomes larger when the SNR increases. Another observation is that if the snapshot number is relatively small, e.g., $N=100$, the two approaches performs nearly the same even at high SNRs. The possible explanation is that the error caused by the limited number of snapshots dominates the approximation error.

\begin{figure}
  \centering
  \includegraphics[width=3.5 in]{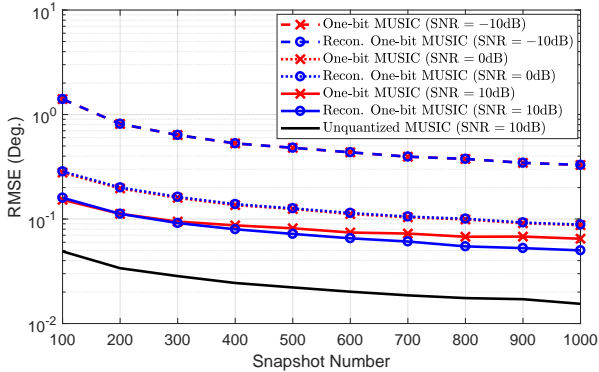}\\
  \caption{RMSE of DOA estimation versus snapshot number.}\vspace{-1em}\label{Fig5}
\end{figure}

In the third example, we consider three cases of ${\text {SNR}} = -10$ dB, $0$ dB and $10$ dB, and vary the number of snapshots $N$ from 100 to 1000 for each case. The resulting RMSEs are shown in Fig. \ref{Fig5}. Again, it is seen that the for mildly low SNRs (e.g., $<0$ dB), the one-bit MUSIC algorithm performs the same as the one-bit MUSIC with covariance matrix reconstruction, regardless of the snapshot number. However, for high SNR and large number of snapshots, performance gap has been caused by approximation error. Nevertheless, the difference would be ignorable in practical applications.

\begin{figure}
  \centering
  \includegraphics[width=3.5 in]{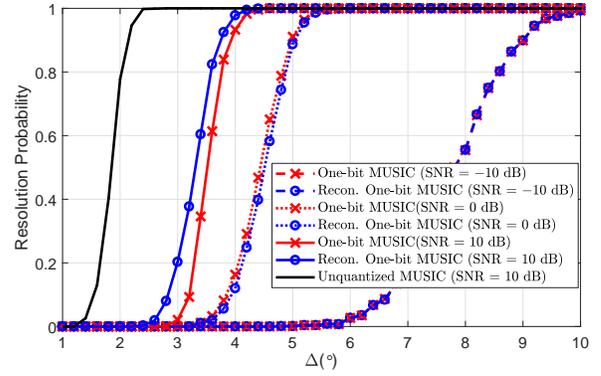}\\
  \caption{Resolution probability versus angular separation.}\vspace{-1em}\label{Fig6}
\end{figure}

Finally, we examine the resolution probability of the one-bit MUSIC algorithm. In particular, we assume the DOAs are $-10^\circ$ and $-10^\circ+\Delta$, where $\Delta \in [1^\circ, 10^\circ]$ denotes the angular separation. In our simulation, if the biases of the two DOA estimates are both less than $\frac{1}{2}\Delta$, it is said to be successfully resolved. Fig. \ref{Fig6} shows the resulting resolution probability versus angular separation. It is seen that only at high SNRs can the extra reconstruction provide certain improvement.

\section{Conclusion}
We have shown by both analysis and simulations that the covariance matrix of one-bit measurements can be straightforwardly utilized as the unquantized covariance matrix to obtain the signal and noise subspaces for DOA estimation, especially at relatively low SNRs. Even though the approximation error become relatively large at high SNRs, it is in general acceptable in practical applications. The findings in this work is helpful to reduce the the system hardware cost and power consumption, and simplify the real-time implementation.

\end{document}